\documentclass[seceq]{ptptex}
\usepackage{wrapft}
\usepackage{graphicx}
\newcommand{\wtilde}[1]{\widetilde{#1}} %%

\newcommand{\lal}{\langle\!\langle}
\newcommand{\rar}{\rangle\!\rangle}
\newcommand{\wb}[1]{\overline{#1}}
\def\beq{\begin{eqnarray}}
\def\eeq{\end{eqnarray}}
\def\bsub{\begin{subequations}}
\def\esub{\end{subequations}}
\def\b{\begin{equation}}
\newcommand\nn{\nonumber \\}

\newcommand{\ldk}{\left[}

\newcommand{\rdk}{\right]}

\title{%        %You can use \\ for explicit line-break
Quark-Hadron Phase-Transition in an Extended NJL Model \\
with Scalar-Vector Interaction}
\subtitle{Finite Temperature and Baryon Chemical Potential Case
}    %use this when you want a subtitle

\author{%       %Use \sc for the family name
Tong-Gyu {\sc Lee},$^{1}$
Yasuhiko {\sc Tsue},$^{2}$ 
Jo\~ao da {\sc Provid\^encia}$^{3}$,\\
Constan\c{c}a {\sc Provid\^encia}$^{3}$
and Masatoshi {\sc Yamamura}$^{4}$ 
}

\inst{%         %Affiliation, neglected when [addenda] or [errata]
$^{1}$Graduate School of Integrated Arts and Sciences, Kochi University, Kochi 780-8520, 
Japan\\
$^{2}$Physics Division, Faculty of Science, Kochi University, Kochi 780-8520, 
Japan\\
$^{3}$Departamento de F\'{i}sica, Universidade de Coimbra, 3004-516 Coimbra, 
Portugal\\
$^4$Department of Pure and Applied Physics, 
Faculty of Engineering Science,\\
Kansai University, Suita 564-8680, Japan
}
\recdate{%      %Editorial Office will fill in this.
\today
}

\abst{%this abstract is neglected when [addenda] or [errata]
The quark-hadron phase transition at finite temperature and baryon chemical potential is investigated in an extended NJL model with scalar-vector eight-point interaction by comparing the pressure of symmetric nuclear matter with that of the quark matter.
As a result, the extended NJL phase diagram is obtained in the temperature-baryon chemical potential plane and the effects of the scalar-vector coupling constant $G_{sv}^q$ on the chiral phase transition are summarized.
In our investigation, the quark-hadron phase transition occurs after the chiral phase transition in the nuclear matter.
It is shown that a quarkyonic-like phase in which the chiral symmetry is restored but the elementary excitation modes are nucleonic appears just before deconfinement in this model. 
}

\begin{document}

%make title
\maketitle

%%%%%%%%%%%%%%%%%%%%%%%%%%%%%
%%          					%%
%%		Introduction		%%
%%          					%%
%%%%%%%%%%%%%%%%%%%%%%%%%%%%%
\section{Introduction}\label{sec:1}
% phase diagram
To understand the strongly interacting many-particle systems governed by Quantum Chromodynamics (QCD) under extreme conditions at high temperature and/or at high density is one of the most fascinating subjects in modern theoretical physics.
Above all, theoretical studies of the phase transition between hadronic and quark-gluon matters and/or the phase diagram on the temperature-chemical potential plane for quark-hadron many-body systems at finite temperature and density are the most recent interests.
In these extremely hot and/or dense environment for quark-hadron systems, there may exist various possible phases with rich symmetry breaking pattern\cite{FH}.
%such as hadronic or quark-gluon phase, chiral symmetric or broken phase and two flavor color-superconducting (2CS, uCS, dCS) or color-flavor locked (CFL) phase and so forth\cite{FH}. 
The extremely high temperature system which is reproduced experimentally by the relativistic heavy ion collisions (RHIC) has been examined theoretically by the 
first principle lattice calculations.
In the finite density system, however, the lattice QCD simulation is not straightforwardly feasible due to the so-called sign problem, namely, it is difficult to understand directly from QCD at finite density.
Thus, the effective model based on QCD can be a useful tool to deal with finite density system.
By using the various effective models, the chiral phase transition has been often investigated at finite temperature and density.
However, it is still difficult to derive the definite results on the quark-hadron phase transition due to the quark confinement on the hadron side.

% NJL
The Nambu-Jona-Lasinio (NJL) model\cite{NJL} is one of the useful effective models of QCD.
This model provides wide information for hadronic systems based on the dynamical chiral symmetry breaking and its restoration\cite{Review-NJL}. 
In considering the nuclear matter, a four-point interaction term which is a characteristic one in the NJL Lagrangian effectively comes out of a string model approach\cite{BPP}, so that we adopt a NJL-type model with the four-point interaction as a model for nuclear matter in this paper.
On the other hand, 
the NJL model was originally introduced for nucleonic/hadronic degrees of freedom\cite{KBKM, Buballa96}.
Some of similar models describe a nucleon as a dynamical quark-diquark bound-state\cite{BT}.
However, the NJL model is mostly used for quark degrees of freedom in the modern form\cite{Buballa05}.
When dealing with the symmetric nuclear matter, it is necessary to reproduce the property of nuclear saturation as the Walecka model\cite{Walecka} has succeeded in describing phenomenologically the saturation property of symmetric nuclear matter without chiral symmetry in which the nucleon is treated as a fundamental particle, not as a composite one.
In the original NJL model with chiral symmetry, however, if the nucleon field is treated as not a composite but as a fundamental fermion field, it is unable to reproduce the nuclear matter saturation property.
Here, it has been observed that the nuclear saturation property is well reproduced by introducing the scalar-vector and isoscalar-vector eight-point interaction in the original NJL model\cite{KBKM}, the nucleon being then a fundamental fermion.
Thus, it is possible to consider a NJL-type model in which the nuclear saturation property is satisfied, as one of the possible models of nuclear matter\cite{MPPM, TPPY}.
For this reason, in this paper, the NJL-type model for nuclear matter with rather reasonable nuclear saturation properties is adopted %.
in which although the NJL-type model for nucleon contains a conceptual problem that an artificial Goldstone mode appears, we regard a meson, as well as a nucleon, as a fundamental particle and do not treat the pion-like excitation here with the same treatment as some works\cite{KBKM, Buballa96,MPPM, TPPY}.
%we regard a meson, as well as a nucleon, as a fundamental particle and ignore the pion-like excitation with the same treatment as some works\cite{KBKM, Buballa96,MPPM, TPPY}.
%

%objective
In this paper, with the aim of extending our previous work\cite{TPPY} to finite temperature and chemical potential case,
we investigate the quark-hadron phase transition and draw the phase diagram between the symmetric nuclear matter and free quark phase without diquark correlation.
Then, for nuclear matter, we adopt the extended NJL model with the scalar-vector eight-point interaction and treat the nucleon field as a fundamental 
fermion field with $N_c=1$ in which $N_c$ is the number of colors.
As for quark matter, we adopt the extended NJL model with $N_c=3$ and treat the quark field as a fundamental fermion field.
%T=0
In the zero-temperature results in Ref.~\citen{TPPY}, it is found that a first-order quark-hadron phase transition is obtained at finite density and the quark-hadron phase transition occurs after the chiral symmetry restoration in nuclear matter.

%organization
This paper is organized as follows:
In the next section, we briefly recapitulate the extended NJL model at finite temperature and baryon chemical potential for nuclear and quark matters following Ref.~\citen{TPPY}.
In $\S$\ref{sec:3}, the numerical results are given and the quark-hadron phase transition is described in this model.
In $\S$\ref{sec:4}, the extended NJL phase diagram with scalar-vector eight-point interaction at finite temperature and baryon chemical potential is presented.
Also, the dependence of the scalar-vector coupling constant on the phase diagram is shown.
The last section is devoted to a summary and concluding remarks.

%%%%%%%%%%%%%%%%%%%%%%%%%%%%%%%%%%%%%%
%%          							%%
%%		Recapitulation of ENJL		%%
%%          							%%
%%%%%%%%%%%%%%%%%%%%%%%%%%%%%%%%%%%%%%
\section{Brief recapitulation of the extended NJL model for nuclear and quark matters at finite temperature and density}\label{sec:2} 

In this section, following Ref.\citen{TPPY}, the same NJL-type models including the scalar-vector eight-point interaction are given for nuclear and quark matters at finite temperature and density, the model parameters and the number of colors being different.

%%%%%%%%%%%%%%%%%%%%%%%%%%%%%%%%%%%%%%%%%%%%%%%%%%%%%%%%
%%          	  										%%
%%	     Lagrangian density, Gap equation and Pressure		%%
%%          		 									%%
%%%%%%%%%%%%%%%%%%%%%%%%%%%%%%%%%%%%%%%%%%%%%%%%%%%%%%%%
\subsection{Lagrangian density, gap equation and pressure}\label{subsec:2-1}
Let us start with the following Lagrangian density for nuclear and quark matters: 
\begin{eqnarray}\label{eq:2-1}
{\cal L}_i
&=& {\wb \psi}_i i\gamma^{\mu}\partial_{\mu} \psi_i
+ G_s^i \ldk ({\wb \psi}_i{\psi}_i)^2+({\wb \psi}_i i\gamma_5{\mib \tau}\psi_i)^2 \rdk \nn
&&
- G_v^i ({\wb \psi}_i\gamma^{\mu}\psi_i)({\wb \psi}_i\gamma_{\mu}\psi_i) \nn
&&
- G_{sv}^i \ldk ( {\wb \psi}_i{\psi}_i)^2 + ({\wb \psi}_i i\gamma_5{\mib \tau}\psi_i)^2 \rdk ({\wb \psi}_i\gamma^{\mu}\psi_i)({\wb \psi}_i\gamma_{\mu}\psi_i) \ ,
\end{eqnarray}
where the subscript/superscript $i$ denotes nothing but an index which represents the case of nuclear matter ($i=N$) or quark matter ($i=q$).
Here, $\psi_i$ represents fermion field, that is,  $\psi_N$ is nucleon field and $\psi_q$ is quark field, and $\mib \tau$ are the Pauli matrices in isospin space.
The first two terms are the original NJL model Lagrangian density.
The third term is a vector-vector repulsive term with $G_v^i$.
The last term is a scalar-vector and isoscalar-vector coupling term with $G_{sv}^i$.
Parameters $G_v^i$ and $G_{sv}^i$ represent a coupling constant of four-point vector-vector interaction and that of eight-point scalar-vector interaction, respectively.
As is well known, the pure NJL interaction alone is not enough to reproduce the property of nuclear saturation in the original NJL model.
Thus, we introduce the last two terms in Eq.~(\ref{eq:2-1}) so as to reproduce the nuclear matter saturation properties\cite{KBKM}.
Then, in this paper, this model is called an extended NJL model. 
This model is nonrenormalizable, so that we adopt a three-momentum cutoff scheme in which the cutoff parameter $\Lambda_{i}$ is introduced.

% MFA
Under the mean field approximation%
\footnote{We make a replacement such as ${\wb \psi}\Gamma \psi \rightarrow \lal {\wb \psi}\Gamma \psi \rar + ({\wb \psi}\Gamma \psi -\lal {\wb \psi}\Gamma \psi \rar)$ where ${\wb \psi} \Gamma \psi$ is bi-linear quantities in the fermion fields and $\Gamma$ are matrices in Dirac, flavor and color space.
The symbol $\lal \cdots \rar$ denotes the expectation value at finite temperature or thermal average.
Here, the fluctuation, ${\wb \psi}\Gamma \psi -\lal {\wb \psi}\Gamma \psi \rar$, is linearized. Namely, we consider two non-vanishing terms: $\lal {\wb \psi}_i\psi_i \rar \neq 0$ and $\lal {\wb \psi}_i\gamma^0\psi_i \rar \neq 0$.
$(\rho_{i} \equiv \lal {\psi}_i^{\dagger}\psi_i \rar = \lal {\wb \psi}_i\gamma^0\psi_i \rar$ where $\rho_{i}$ represents the fermion number density.)
},
the Lagrangian density ${\cal L}_i^{MF}$ and the Hamiltonian density ${\cal H}_i^{MF}$ are obtained as
\begin{eqnarray}
&&{\cal L}_i^{MF} = {\wb \psi}_i(i\gamma^{\mu}\partial_{\mu}-m_i)\psi
-{\wtilde \mu}_i{\wb \psi}_i\gamma^0\psi_i + C_i \ , \nonumber \\
&&{\cal H}_i^{MF} = -i{\wb \psi}_i{\mib \gamma}\cdot \nabla\psi_i
+m_i{\wb \psi}_i\psi_i
+{\wtilde \mu}_i{\wb \psi}_i\gamma^0\psi_i -C_i \ , \nonumber\\
&& \qquad
C_i \equiv -G_s^i\lal{\wb \psi}_i\psi_i\rar^2+G_v^i\lal{\wb \psi}_i\gamma^0\psi_i\rar^2
+3G_{sv}^i\lal{\wb \psi}_i\psi_i\rar^2\lal{\wb \psi}_i\gamma^0\psi_i\rar^2 \ ,  \label{eq:2-2}
\end{eqnarray}
where
\begin{eqnarray}
m_i &=& -2 \ldk G_s^i - G_{sv}^i \lal{\wb \psi}_i\gamma^0\psi_i\rar^2 \rdk \lal{\wb \psi}_i\psi_i\rar \ , \label{eq:2-3}\\
{\wtilde \mu}_i &=& 2 \ldk G_v^i + G_{sv}^i \lal{\wb \psi}_i\psi_i\rar^2 \rdk \lal{\wb \psi}_i\gamma^0\psi_i\rar \ \label{eq:2-4}
\end{eqnarray}
for nuclear matter ($i=N$) and quark matter ($i=q$), respectively. Here, the symbol $\lal \cdots \rar$ denotes the finite-temperature expectation value which represents thermal average.
% ($\lal {\cal O} \rar = \mathsf{Tr} {\cal O} e^{-\beta ({\cal H}^{MF}_i -\mu_{i}{\cal N}_i)} / \mathsf{Tr} e^{-\beta ({\cal H}^{MF}_i -\mu_{i}{\cal N}_i)}$). 

% chemical potential
In addition, we introduce the chemical potential $\mu_i$ to deal with a finite density system:
\begin{eqnarray}
&&{\cal H}_i'
= {\cal H}_i^{MF} - \mu_i\psi_i^\dagger \psi_i \nonumber\\
&&\quad= -i{\wb \psi}_i{\mib \gamma}\cdot \nabla\psi_i+m_i{\wb \psi}_i\psi_i
-{\mu}_i^r{\wb \psi}_i\gamma^0\psi_i -C_i \ , \label{eq:2-5}
\end{eqnarray}
where $\mu_i^r$ is the effective chemical potential:
\begin{equation}
\mu_i^r=\mu_i-{\wtilde \mu}_i
=\mu_i- 2 \ldk G_v^i + G_{sv}^i\lal{\wb \psi}_i\psi_i\rar^2 \rdk \lal{\wb \psi}_i\gamma^0\psi_i\rar \ .  \label{eq:2-6}
\end{equation}
Here, the expectation values at finite temperature is given as
\begin{eqnarray}
&&\lal {\wb \psi}_i\psi_i \rar 
=\nu_i\int \frac{d^3{\mib p}}{(2\pi)^3}\frac{m_i}{\sqrt{{\mib p}^2+m_i^2}}
(n_+^i -n_-^i)\ , \label{eq:2-7}\\
&&\lal {\wb \psi}_i\gamma^0\psi_i \rar 
=\nu_i\int \frac{d^3{\mib p}}{(2\pi)^3}
(n_+^i + n_-^i -1)\  \label{eq:2-8}
\end{eqnarray}
with
\begin{eqnarray}
&&\nu_i = 2 N_f^i N_c^i \ ,
\label{eq:2-9} \\
&&n_{\pm}^i = \left[ e^{\beta(\pm\sqrt{{\bf p}^2+m_i^2}-\mu_i^r)}+1 \right]^{-1} \ ,
\label{eq:2-10}
\end{eqnarray}
where $\nu_i$ is the degeneracy factor in which $N_c^i$ and $N_f^i$ represent the numbers of color and flavor, and $n_{\pm}^i$ are the fermion number distribution functions with $\beta=1/T$ in which $T$ is temperature.
Here, we have eliminated the contribution of the occupied negative energy states from the nucleon and/or quark number density itself in Eq.~(\ref{eq:2-8}).
As a result, we obtain a self-consistent set of equations, Eqs.~(\ref{eq:2-3}) and (\ref{eq:2-6})-(\ref{eq:2-8}) with (\ref{eq:2-10}).
A self-consistent equation for $m_i$ in Eq.~(\ref{eq:2-3}) is noting but the so-called gap equation in the BCS theory.

% pressure
The thermodynamic potential density $\omega_{i}$ for nuclear and quark matters are defined as
\begin{eqnarray}
\omega_{i} &=& \lal {\cal{H}}^{MF}_i \rar - \mu_i \lal {\cal{N}}_i \rar -\frac{1}{\beta} \lal S_i \rar \ , \label{eq:2-11}
\end{eqnarray}
where
\begin{eqnarray}
&&\lal {\cal{H}}^{MF}_i \rar = \lal {\wb \psi}_i ({\mib \gamma}\cdot{\mib p}) \psi_i \rar - G_{s}^i \lal {\wb \psi}_i\psi_i \rar^2 \nn
&&\qquad\qquad\qquad\qquad\quad
+ G_v^i \lal {\wb \psi}_i\gamma^0\psi_i \rar^2 + G_{sv}^i \lal {\wb \psi}_i\psi_i \rar^2 \lal {\wb \psi}_i\gamma^0\psi_i \rar^2 , \label{eq:2-12} \\
&&\lal {\cal{N}}_i \rar = \lal {\wb \psi}_i\gamma^0\psi_i \rar \ , \label{eq:2-13} \\
&&\lal S_i \rar  = -\nu_i\int \frac{d^3{\mib p}}{(2\pi)^3} 
\bigm[ {n_+^i \ln n_+^i} + {(1-n_+^i) \ln (1-n_+^i)} \nn
&&\qquad\qquad\qquad\qquad\qquad\qquad
+ {n_-^i \ln n_-^i} + {(1-n_-^i) \ln (1-n_-^i)} \bigm] \ , \label{eq:2-14}
\end{eqnarray}
and
\begin{eqnarray}
\lal {\wb \psi}_i ({\mib \gamma}\cdot{\mib p}) \psi_i \rar
= \nu_i\int \frac{d^3{\mib p}}{(2\pi)^3} \frac{{\mib p}^2}{\sqrt{{\mib p}^2+m_i^2}} (n_+^i -n_-^i) \ . \label{eq:2-15}
\end{eqnarray}
Incidentally, we obtain the gap equation in Eq.~(\ref{eq:2-3}) and the fermion number distribution functions in Eq.~(\ref{eq:2-10}) again by minimizing $\omega_i$ with respect to $m_i$ and $n_{\pm}^i$. 
From this thermodynamic potential density, the pressure of nuclear matter ($i=N$) and that of quark matter ($i=q$) are given as
\begin{eqnarray}
p_{i} &=& - \ldk \lal {\cal{H}}^{MF}_i\rar(\rho_i) - \lal{\cal{H}}^{MF}_i\rar(\rho_i=0) \rdk + \mu_i \lal {\cal{N}}_i \rar + \frac{1}{\beta} \lal S_i \rar \ , \label{eq:2-16}
\end{eqnarray}
where we subtracted the zero-density expectation value of the mean field Hamiltonian in the vacuum.
This subtraction method is the same as that used in the expression for energy density which is given later.
We will discuss the determination of the realized phase by comparing the pressure of nuclear matter and of quark matter.

%%%%%%%%%%%%%%%%%%%%%%%%%%%%%
%%          					%%
%%	    Model-parameters	%%
%%          					%%
%%%%%%%%%%%%%%%%%%%%%%%%%%%%%
\subsection{Model-parameters}\label{subsec:2-2}
For nuclear matter, the extended NJL model has four parameters: $G_{s}^{N}, G_{v}^{N}, G_{sv}^{N}$ and $\Lambda_{N}$. 
These parameters are determined by four conditions at zero temperature: the nucleon mass in vacuum, $m_{N}(\rho_N$=$0) = 939$ MeV, 
the normal nuclear density, $\rho^0_N=0.17 $/fm$^{3}$, the nucleon mass at normal nuclear density, $m_{N}(\rho^0_N) = 0.6{m_N}{(\rho_N=0)}$, 
and the saturation properties of nuclear matter, $W_N(\rho^0_N) = -15$ MeV.
At zero temperature, the fermion number distribution function $n^{\pm}_{i}$
in Eq.~(\ref{eq:2-10}) reduces to the Heaviside step function, $n^+_i = \theta(\mu^r_i - \sqrt{{\mib p}^2 + m^2_i})$, and $n_i^- =1$.
Thus, in the nuclear matter case, the gap equation at zero temperature is expressed as
\begin{eqnarray}
&&m_{N} = -2G_s^N \ldk 1 - \frac{G^N_{sv}}{G^N_s} \rho^2_N \rdk  \langle {\wb \psi}_N \psi_N \rangle \ , \\ \label{eq:2-17}
&&\langle {\wb \psi}_N \psi_N \rangle = -\frac{\nu_N m_N}{2\pi^2} \int_{p^N_F}^{\Lambda_N} d|{\mib p}| \frac{{\mib p}^2}{\sqrt{{\mib p}^2+m_N^2}} \ , \label{eq:2-18}
\end{eqnarray}
where
\begin{eqnarray}
&&\rho_{N} = \frac{\nu_N}{6\pi^2}{p^{N}_{F}}^3 \ , \\ \label{eq:2-19}
&&{p^{N}_{F}}=\sqrt{{\mu^r_N}^2-m_N^2} \ , \label{eq:2-20} 
\end{eqnarray}
and $\nu_N=2 N_f^N N_c^N$ with $N_f^N=2$ and $N_c^N=1$.
The symbol $\langle\cdots \rangle$ denotes the expectation value at zero temperature.
The energy density per single nucleon at finite baryon density and zero temperature is evaluated as
\begin{eqnarray}
&&W_N (\rho_N) = \frac{\langle{\cal{H}}^{MF}_N\rangle(\rho_N) - \langle{\cal{H}}^{MF}_N\rangle(\rho_N=0)}{\rho_N} - m_N(\rho_N=0) \ , \label{eq:2-21}
\end{eqnarray}
where
\begin{eqnarray}
&&\langle{\cal{H}}^{MF}_N\rangle(\rho_N)=\langle {\wb \psi}_N ({\mib \gamma}\cdot{\mib p}) \psi_N \rangle -G_s^N \ldk 1-\frac{G^N_{sv}}{G^N_s} \rho_N^2 \rdk \langle {\wb \psi}_i \psi_i \rangle^2 + G^N_v \rho_N^2 \ , \\ \label{eq:2-22}
&&\langle {\wb \psi}_N ({\mib \gamma}\cdot{\mib p}) \psi_N \rangle=-\frac{\nu_N}{2\pi^2} \int_{p^N_F}^{\Lambda_N} d|{\mib p}| \frac{|{\mib p}|^4}{\sqrt{{\mib p}^2+m_N^2}} \ . \label{eq:2-23}
\end{eqnarray}
Here, we take the form of subtracting vacuum value for $\langle{\cal{H}}^{MF}_N\rangle$ in Eq.~(\ref{eq:2-21}). 
The values of model-parameters for nuclear matter are summarized in Table I.

%
% nuclear-table
\begin{table}[t]
\caption{The parameter set for nuclear matter ($i=N$).}
\label{tb:1}
\begin{center}
\begin{tabular}{c|c|c|c} 
$\quad\Lambda_N$[MeV]$\quad$
& $\quad G_s^N\Lambda_N^2 \quad$
& $\quad G_v^N\Lambda_N^2 \quad$
& $\quad$ $G_{sv}^N\Lambda_N^8$ $\quad$ \\
\hline \hline 
$\quad 377.8 \quad$
& $\quad 19.2596 \quad$
& $\quad 17.9824 \quad$
& $\quad -1069.89\quad \quad$ 
\end{tabular}
\end{center}
\end{table}
%
%
% quark-table
\begin{table}[t]
\caption{The parameter set for quark matter ($i=q$).}
\label{tb:2}
\begin{center}
\begin{tabular}{c|c|c|c}  
$\quad \Lambda_q$[MeV]$\quad$
& $\quad G_s^q\Lambda_q^2 \quad$
& $\quad G_v^q\Lambda_q^2 \quad$
& $\quad$ $G_{sv}^q\Lambda_q^8$ $\quad$
\\
\hline \hline
$\quad$653.961$\quad$
& $\quad$2.13922$\quad$
& $\quad$0$\quad$
& $\quad$free$\quad$
\end{tabular}
\end{center}
%\end{wraptable}
\end{table}
%

% model-characteristics
As may be seen in this table, the momentum cutoff $\Lambda_N$ is rather small.
It is possible, however, that the cutoff similarly increases with density by considering a chemical potential dependent cutoff\cite{add12}.
Under the parameters of Table \ref{tb:1}, a rather reasonable value of the incompressibility is obtained numerically as $K\approx260$ MeV
in which the incompressibility of nuclear matter at normal nuclear density is evaluated as
$K = 9\rho^0_N {d^2 W_N(\rho_N)}/{d \rho^2_{N}} |_{\rho=\rho^0_N}$.
Although we fix the value $m_N(\rho_N^0)$ in this paper,
the incompressibility $K$ at normal nuclear density is capable of becoming an input parameter 
instead of the nucleon mass $m_N(\rho_N^0)$ at normal nuclear density since $m_N(\rho_N^0)$ has influence on $K$.

% quark
For the quark matter, there are three parameters, that is, $G_s^q, G_{sv}^q$ and $\Lambda_q$, in the extended NJL model with scalar-vector eight-point interaction.
Here, we put $G_v^q=0$ % for the quark matter since $G_v^i$ and $G_{sv}^i$ are introduced so as to reproduce the nuclear matter saturation properties for the nuclear matter.
%because the effects of the vector coupling $G_v^q$ is self-evident\cite{Gv}. 
since the effects of the vector coupling $G_v^q$ is %self-evident$^{12)}$.
well-kown\cite{Gv}.
%However, we introduced the parameter $G_{sv}^q$. The reason why is that $G_{sv}^q$ has influence on the chiral phase transition at finite density and temperature.
%On the other hand, we introduced the parameter $G_{sv}^q$ since $G_{sv}^q$ has influence on the chiral phase transition at finite density and temperature.
On the other hand, we introduced the parameter $G_{sv}^q$ because it has influence on the chiral phase transition at finite density and temperature.
The scalar-vector attractive interaction %add
with $G_{sv}^q$ makes the chiral condensate strength stronger.
Namely, the chiral phase transition point is pushed to higher-density side with increasing $G_{sv}^q$%.
%, because the scalar coupling $G_{s}^q$ is regarded as a density-dependent coupling $G_{s}^q(\rho_q)$ effectively by introducing $G_{sv}^q$.
, in which the scalar coupling $G_{s}^q$ is regarded as a density-dependent coupling $G_{s}^q(\rho_q)$ effectively by introducing $G_{sv}^q$.
In this paper, we treat $G_{sv}^q$ as a chiral phase transition tuning parameter.
The parameters $G_s^q$ and $\Lambda_q$ are determined by two conditions: the vacuum value for the dynamical quark mass, $m_q=313$ MeV, 
and the pion decay constant, $f_{\pi}=93$ MeV.
For the value of $G_{sv}^q$, there is no criterion to determine it in this stage.
Thus, we treat  $G_{sv}^q$ as a free parameter.
The physical quantities of quark matter case are obtained from the corresponding ones for the nuclear matter case with $i=N \rightarrow i=q$.
Namely, $m_{N} \rightarrow m_{q}$, $\mu^{r}_{N} \rightarrow \mu^{r}_{q}$ and $\nu_{N}=2N_f^N N_c^N \rightarrow \nu_q=2N_f^q N_c^q$ with $N_f^q=2, N_c^q=3$.
The value of model-parameters for quark matter are summarized in Table II.

%%%%%%%%%%%%%%%%%%%%%%%%%%%%%
%%	  					%%
%%	   Numerical Analysis		%%
%%          					%%
%%%%%%%%%%%%%%%%%%%%%%%%%%%%%
\section{Numerical results}\label{sec:3}
%
%%%%%%%%%%%%%%%%%%%%%%%%%%%%%%%%%%%%%%%%%%%%%%%%%%%%%%%%
%%	  											%%
%%		   Gap-solutions and chiral phase transition		%%
%%          											%%
%%%%%%%%%%%%%%%%%%%%%%%%%%%%%%%%%%%%%%%%%%%%%%%%%%%%%%%%
\subsection{Gap-solutions and chiral phase transition}\label{subsec:3-1}
In the quark matter case, the coupling constant for the scalar-vector and isoscalar-vector eight-point interaction, $G^q_{sv}$, is a free parameter as was already mentioned in previous section.
Hence, we put $G^q_{sv}\Lambda^8_q=-68.4$ so as to realize the chiral phase transition in the reasonable point.
Then, we take the parameter $G^q_{sv}$ as $m_q(\rho_q/3=\rho^0_{N})=0.625m_q(\rho_q=0)$\cite{TPPY}. 
The numerical results which will be presented correspond to this parameter set and to the gap solutions 
in Eq.~(\ref{eq:2-3}) with $G^q_{sv}\Lambda^8_q=-68.4$ at finite temperature and quark chemical potential. 
%
% mass, density-Fig.
%\begin{figure}[t]
%\begin{center}
%   \begin{tabular}{ccc}
%     \resizebox{70mm}{!}{\includegraphics{mq-68.4.eps}} &
%     \resizebox{70mm}{!}{\includegraphics{Rq-68.4.eps}} 
%   \end{tabular}
%\caption{Left: The dynamical quark mass $m_q$ as a function of the quark chemical potential $\mu_q$ with $G_{sv}^q\Lambda_q^8=-68.4$ at $T=0, 20$ and $40$ MeV.
%Right: The quark number density divided by normal nuclear density $\rho_q/\rho_N^0$ as a function of the quark chemical potential $\mu_q$ with $G_{sv}^q\Lambda_q^8=-68.4$ at $T=0, 20$ and $40$ MeV. The solid curves and branch lines are stable solutions. The dashed curves are unstable solutions.
%}
%\label{fig:mR-68.4}
%\end{center}
%\end{figure}
%

\begin{figure}[t]
\begin{center}
   \begin{tabular}{ccc}
     \resizebox{70mm}{!}{\includegraphics{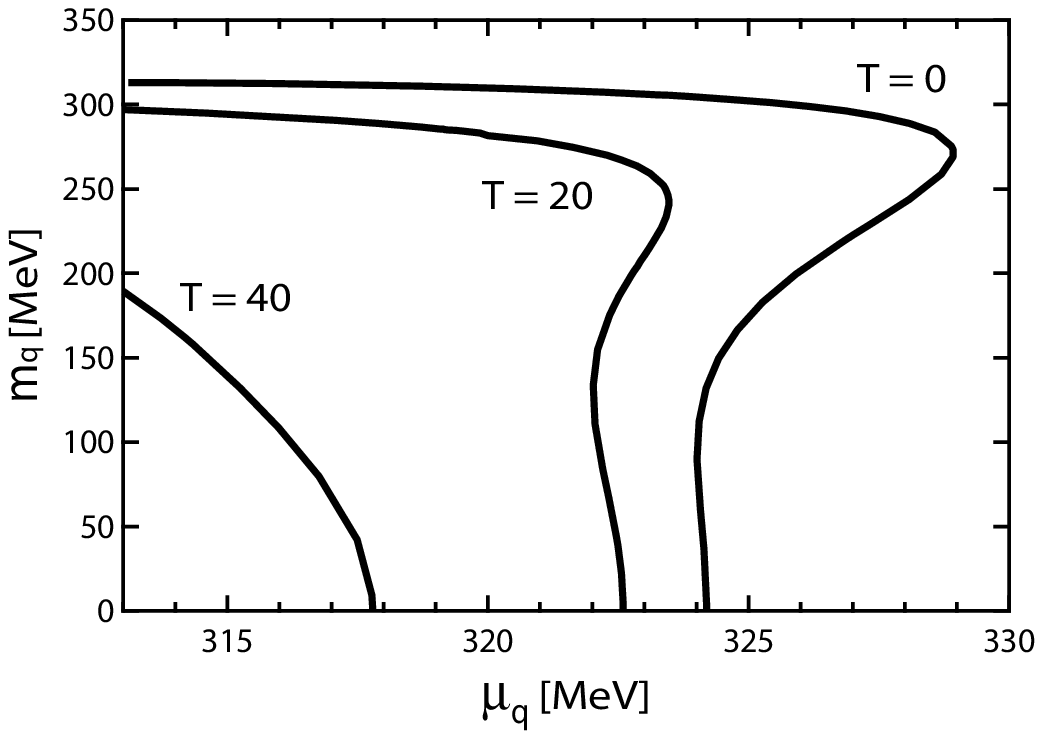}} &
     \resizebox{70mm}{!}{\includegraphics{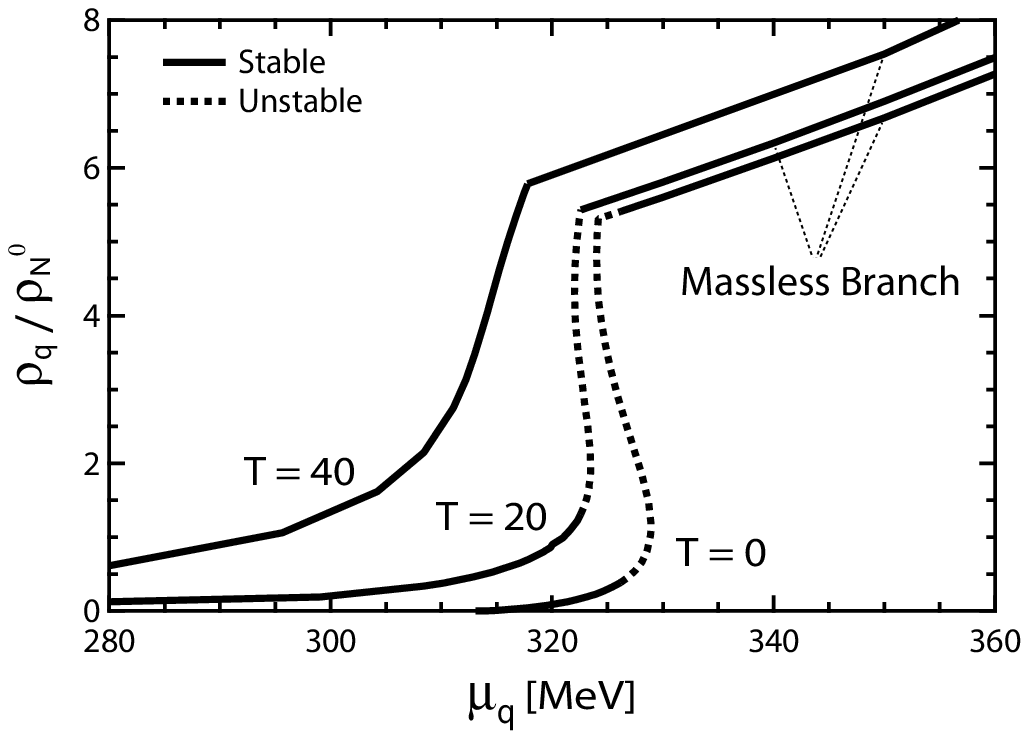}} 
   \end{tabular}
\caption{Left: The dynamical quark mass $m_q$ as a function of the quark chemical potential $\mu_q$ with $G_{sv}^q\Lambda_q^8=-68.4$ at $T=0, 20$ and $40$ MeV.
Right: The quark number density divided by normal nuclear density $\rho_q/\rho_N^0$ as a function of the quark chemical potential $\mu_q$ with $G_{sv}^q\Lambda_q^8=-68.4$ at $T=0, 20$ and $40$ MeV. The solid curves and branch lines are stable solutions. The dashed curves are unstable solutions.
}
\label{fig:mR-68.4}
\end{center}
\end{figure}

% left side
The left side of Fig.~\ref{fig:mR-68.4} shows the dynamical quark mass as a functions of the quark chemical potential at $T=0, 20$ and $40$ MeV.
For zero temperature, the obtained vacuum value for the dynamical quark mass is $\mu_q=313$ MeV.
As may be seen in this figure, the region with the multiple solutions for the gap equation shrinks with increasing temperature.

% right side
The right side of Fig.~\ref{fig:mR-68.4} shows the quark number density as a function of the quark chemical potential at $T=0, 20$ and $40$ MeV.
Here, the quark number density is given as a multiple of normal nuclear density $\rho_N^0$ in the vertical axis.
The relation between the quark number density $\rho_q$ and the quark chemical potential $\mu_q$ is obtained from Eq.~(\ref{eq:2-8}).
The solid curves and branch lines show the stable solutions in this figure.
Here, the solid branch lines represent the solutions of quark number density with massless solution, $\rho_q(m_q$=$0)$.
The dashed curves represent the unphysical region.

\begin{figure}[t]
\begin{center}
\includegraphics[height=5.3cm]{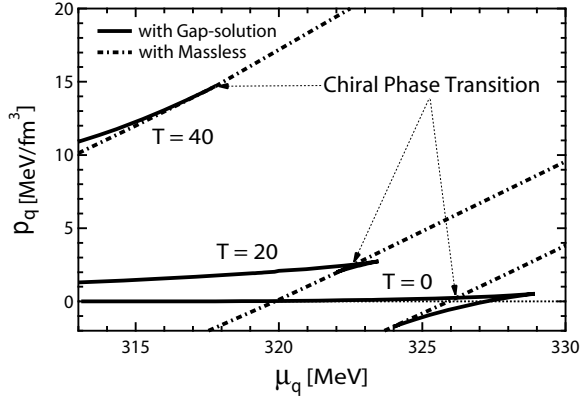}
\caption{The pressure of quark matter $p_q$ as a function of the quark chemical potential 
$\mu_q$ with $G_{sv}^q\Lambda_q^8=-68.4$ at $T=0, 20$ and $40$ MeV. The solid lines represent the pressure with gap-solution, 
while the dash-dotted lines represent the pressure with massless solution. The chiral phase transition at $T=0, 20$ and 40 MeV occur at 
$\mu_q\approx 326, 323$ and $318$ MeV, respectively.
}
\label{fig:Pq-68.4}
\end{center}
\end{figure}

% pressure
The gap equation has multiple solutions in certain region, so that there exists an unphysical region with unstable solutions there.
Thus, we determine the physically realized solution by calculating the pressure $p_q$ in Eq.~(\ref{eq:2-16}) where $i=q$ for quark matter. 
The physically realized solution corresponds to the largest pressure at each temperature.
Figure~\ref{fig:Pq-68.4} shows the pressure of quark matter as a function of the quark chemical potential.

% T=0
In the case of $T=0$ MeV, the lower density solution is realized from $\mu_q=313$ MeV to $\mu_q\approx 326$ MeV in Fig.~\ref{fig:mR-68.4} on the left where the pressure is the largest value.
Above $\mu_q\approx 326$ MeV, however, the massless solution becomes physically realized.
Namely, the chiral phase transition for $T=0$ occurs at $\mu_q\approx 326$ MeV as seen in Fig.~\ref{fig:Pq-68.4}.
In the quark phase, the phase with dynamical quark mass (chiral broken phase) is realized for 
the lower quark chemical potential ($\mu_q<326$ MeV) and the phase with massless quark (chiral symmetric phase) 
for the higher one ($\mu_q>326$ MeV) as may be seen from Fig.~\ref{fig:Pq-68.4}.
In the region from $\rho_q \sim 0.38\rho_N^0$  ($\rho_B \sim 0.13\rho_N^0$ where $\rho_B$ represents the baryon number density) to 
$\rho_q \sim 5.41\rho_N^0$ ($\rho_B \sim 1.80\rho_N^0$) in Fig.~\ref{fig:mR-68.4} on the right, 
a first-order chiral phase transition is realized and the coexistence of quark phases occurs.

% mass, density-Fig.
%
% pressure-Fig.
%

% T=20
In the case of $T=20$ MeV, the low density solution is realized up to $\mu_q\approx 323$ MeV and the massless solution becomes physically realized from $\mu_q\approx 323$ MeV in Fig.~\ref{fig:mR-68.4} on the right.
From Fig.~\ref{fig:Pq-68.4}, it is seen that the chiral phase transition with $T=20$ MeV occurs at $\mu_q\approx 323$ MeV.
In this case, the chiral broken phase is realized at $\mu_q<323$ MeV and the chiral symmetric phase at $\mu_q>323$ MeV.
Here, in the region from $\rho_q \sim 1.30\rho_N^0$  ($\rho_B \sim 0.43\rho_N^0$) to $\rho_q \sim 5.41\rho_N^0$ ($\rho_B \sim 1.80\rho_N^0$) in Fig.~\ref{fig:mR-68.4} on the right, a first-order chiral phase transition is realized and the coexistence of quark phases occurs.

% T=40
In the case of $T=40$ MeV, the density solution with gap-solution is realized in all points (from $\mu_q=0$ to $\mu_q\approx 318$ MeV) 
and the massless solution becomes physically realized from $\mu_q\approx318$ MeV in Fig.~\ref{fig:mR-68.4} on the right.
In Fig.~\ref{fig:Pq-68.4} with $T=40$ MeV, two branches are smoothly connected and at $\mu_q \approx 318$ MeV 
a chiral phase transition occurs, or at $\rho_q \sim 5.78\rho_N^0$ ($\rho_B\sim1.93\rho_N^0$) in Fig.~\ref{fig:mR-68.4} on the right.
In this case, the phase with chiral symmetry breaking is realized for $\mu_q<318$ MeV and the phase with chiral restoration for $\mu_q>318$ MeV.
Unlike in the case of $T=0$ or $20$ MeV, 
at $T=40$ MeV, there is no jump from the lower density solution to massless solution since there is no unstable density solution as 
Fig.~\ref{fig:mR-68.4} on the right with $T=40$ MeV shows. 
Hence, the coexistence of quark phases doesn't occur and the order of the chiral phase transition is not a first-order phase transition.
In case that there exists a jump as in Fig.~\ref{fig:mR-68.4} on the right with $T=0$ or $20$ MeV, 
it is clear that the phase transition is of first order.
However, because there is no jump as in Fig.~\ref{fig:mR-68.4} on the right with $T=40$ MeV  
and the slope of the pressure with respect to quark chemical potential is the same, namely, two branches are smoothly connected as is seen in Fig.~\ref{fig:Pq-68.4}, 
the order of the phase transition at $T=40$ MeV 
is a second order. 

%we are unable to determine the order of the chiral phase transition under this simple discussion.
%Thus, we only determine whether the phase transition is a first-order phase transition in this stage.

% supplementation
Incidentally, this critical density of the chiral phase transition is lower than that of the quark-hadron phase transition as is mentioned below.
Also, the chiral symmetry restoration occurs systematically at the lower density side, so that it has no actual physical consequences and no influence on the curve of the equation of state in the quark phase after the quark-hadron phase transition.

%%%%%%%%%%%%%%%%%%%%%%%%%%%%%%%%%%%%%%
%%	  							%%
%%	Quark-Hadron phase transition		%%
%%          							%%
%%%%%%%%%%%%%%%%%%%%%%%%%%%%%%%%%%%%%%
\subsection{Quark-Hadron phase transition}\label{subsec:3-2}
The main object of this paper is to investigate the quark-hadron phase transition in the extended NJL model with the scalar-vector eight-point interaction at finite temperature and density.
In this subsection, we present a procedure for investigating the phase transition between the nuclear and quark matters and show the numerical results.

% procedure for investigating
For the quark-hadron phase transition, we follow the same approach to determine the physically realized phase 
as it has already been shown in previous subsection.
Namely, we determine the realized phase by comparing the pressure of nuclear matter with that of quark matter at finite temperature and baryon chemical potential.
For this purpose, the condition for chemical equilibrium is demanded as
\begin{eqnarray}
\mu_N (T) = 3\mu_q (T) \ , \label{eq:3-1}
\end{eqnarray}
where $\mu_N$ and $3\mu_q$ indicate the chemical potential per baryon.
By regarding this condition as the one for thermodynamic equilibrium between the hadron and the quark phases, we derive the corresponding condition for the pressure of hadron and quark phases as 
\begin{eqnarray}
p_N (\mu_N, T) =  p_q (3\mu_q, T) \ . \label{eq:3-2}
\end{eqnarray}
From Eq.~(\ref{eq:2-16}), the pressure $p_N$ and $p_q$ can be calculated for nuclear matter and for quark matter, respectively.

% Quark-Hadron(T=0)
\begin{figure}[t]
\begin{center}
   \begin{tabular}{ccc}
     \resizebox{70mm}{!}{\includegraphics{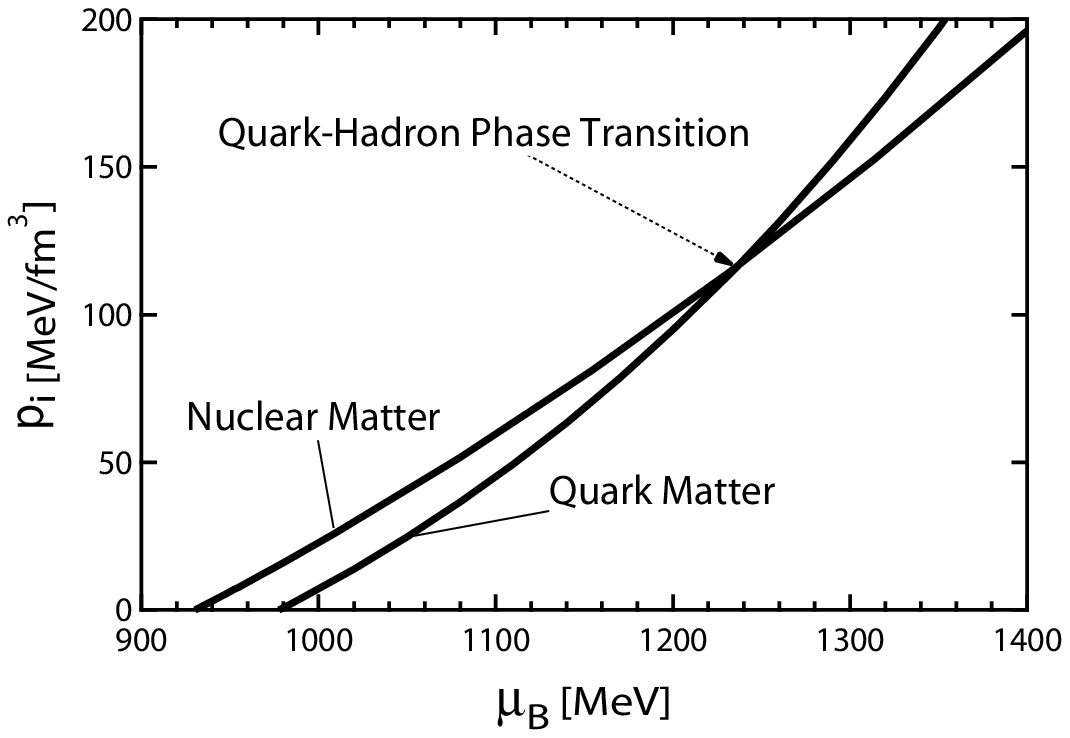}} &
     \resizebox{70mm}{!}{\includegraphics{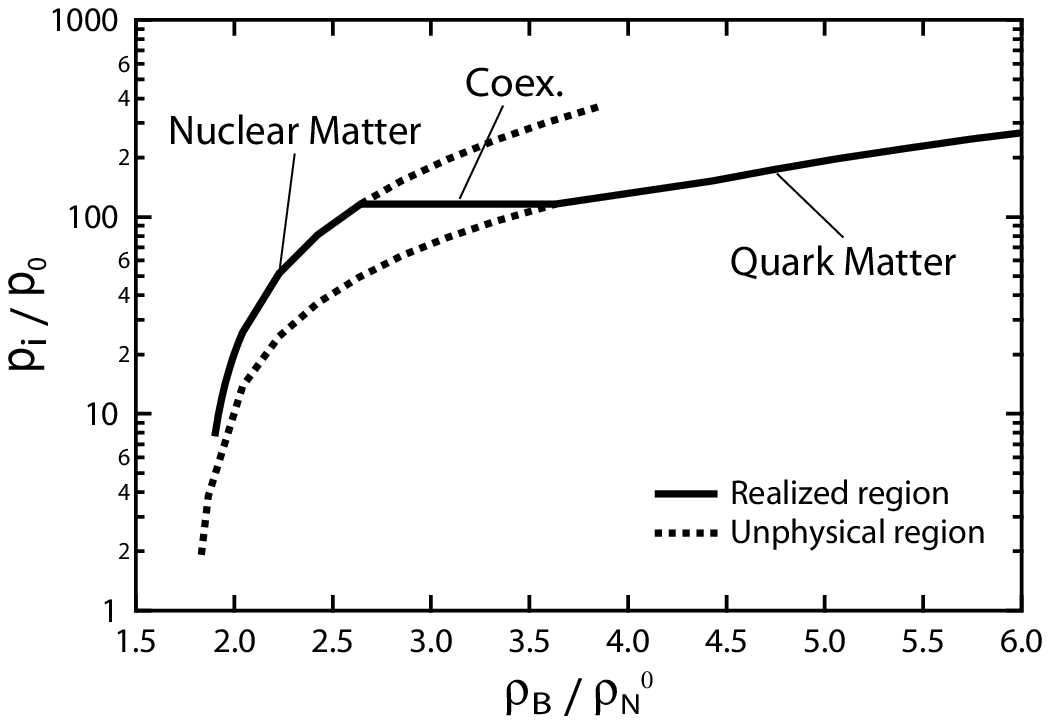}} 
   \end{tabular}
\caption{The quark-hadron phase transition is shown at $T=0$ in the case of $G_{sv}^q\Lambda^8_q=-68.4$.
Left: The pressure of nuclear matter ($i=N$) and of quark matter ($i=q$) as a function of the baryon chemical potential $\mu_B$ ($=\mu_N=3\mu_q$) at $T=0$ MeV.
Right: The pressure $p_i/p_0$ as a function of the baryon number density $\rho_B/\rho_N^0$ at $T=0$ MeV. Here, the vertical axis is shown in logarithmic scale. The pressure is divided by $p_0$ where $p_0=1.0$MeV/fm$^3$. The baryon number density is given in multiples of normal nuclear density $\rho_N^0$. 
}
\label{fig:QH0}
\end{center}
\end{figure}
%
% Left side
Figure~\ref{fig:QH0} on the left shows the pressure of nuclear and of quark matter as a function of the baryon chemical potential 
which is equivalent to the nuclear chemical potential and the triple of the quark chemical potential at $T=0$ in the case of 
$G_{sv}^q\Lambda_q^8=-68.4$.
As may be seen from this figure, there is a crossing point at a certain chemical potential value.
From this crossing point, we can determine the coexistence of nuclear and quark phases.
Then, about $\mu_B\approx1236$ MeV, the quark-hadron phase transition at $T=0$ occurs in the left side of Fig.~\ref{fig:QH0}.
Thus, the nuclear phase (hadron phase) is realized for the smaller baryon chemical potential ($\mu_B=\mu_N<1236$ MeV) 
and the quark phase for the larger chemical potential ($\mu_B=3\mu_q>1236$ MeV).

% Right side
On the other hand, Fig.~\ref{fig:QH0} on the right shows the pressure as a function of the baryon number density at $T=0$ 
where the vertical axis is shown in logarithmic scale. 
This figure is depicted by using the relation of the density and chemical potential shown in Fig.~\ref{fig:mR-68.4} on the right.
It is seen from this figure that the nuclear and quark phases coexist and a first-order quark-hadron phase transition occurs 
in the region from $\rho_N (=\rho_B) \sim 2.64\rho_N^0$ to $\rho_q \sim 10.9\rho_N^0$ ($\rho_B \sim 3.63\rho_N^0$).

% Quark-Hadron(T=20)
\begin{figure}[t]
\begin{center}
   \begin{tabular}{ccc}
     \resizebox{70mm}{!}{\includegraphics{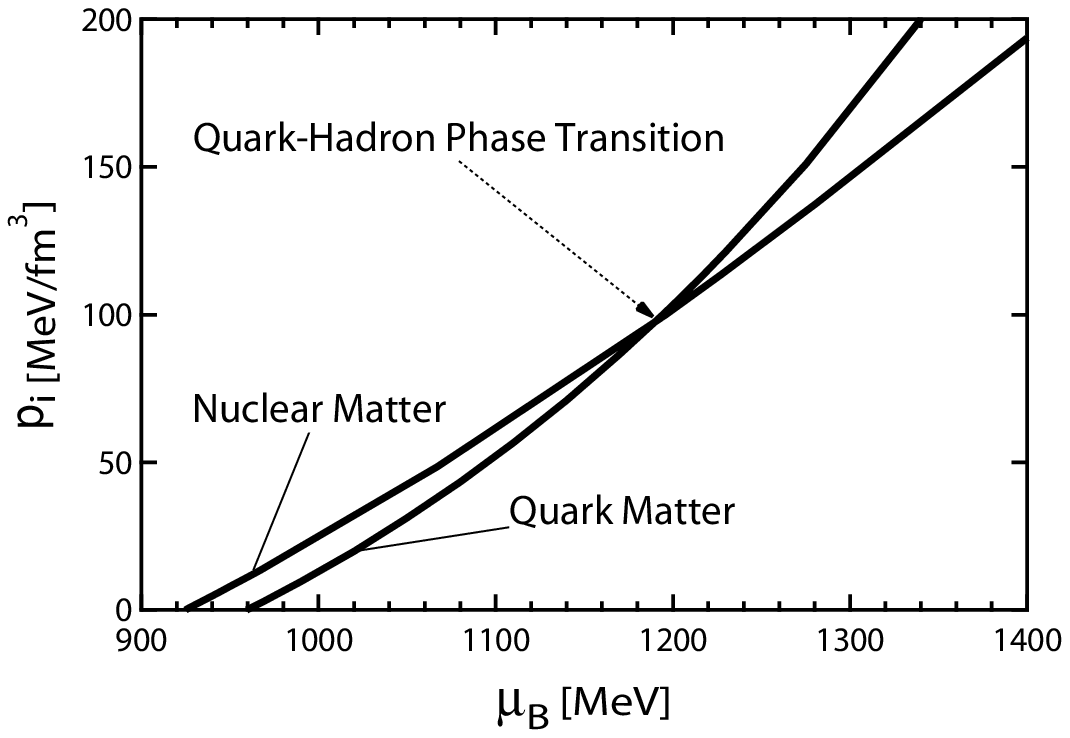}} &
     \resizebox{70mm}{!}{\includegraphics{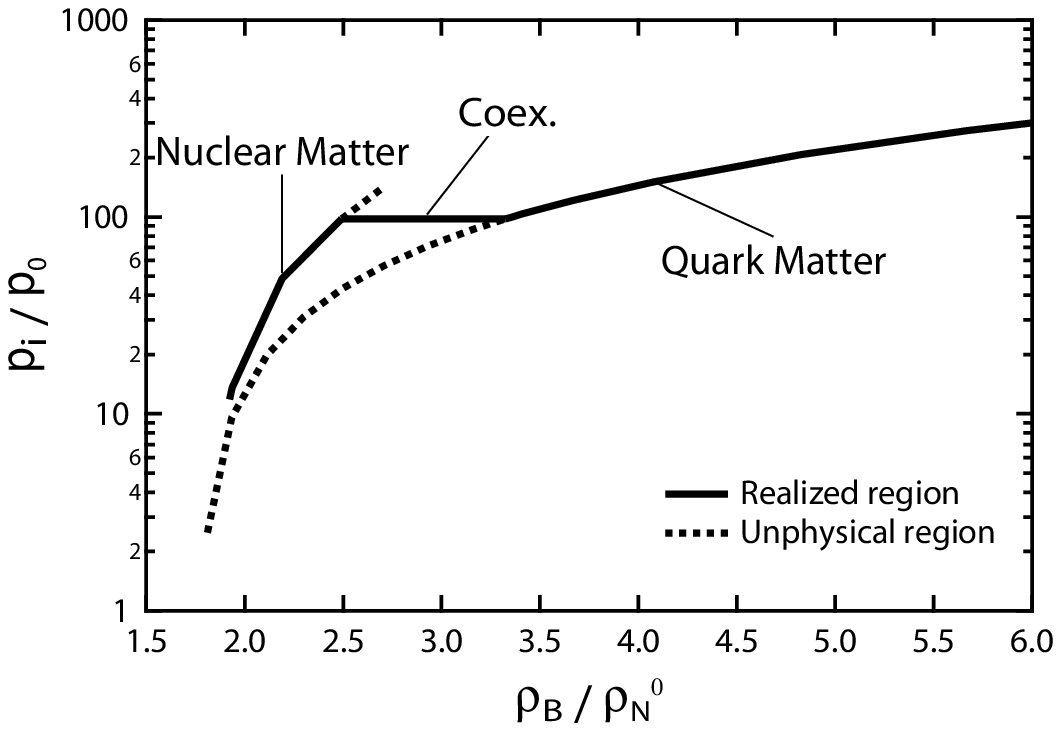}} 
   \end{tabular}
\caption{The quark-hadron phase transition is shown at $T=20$ MeV with $G_{sv}^q\Lambda^8_q=-68.4$.
Left: The pressure of nuclear matter and of quark matter as a function of the baryon chemical potential $\mu_B$ at $T=20$ MeV.
Right: The pressure presented on a logarithmic scale $p_i/p_0$ as a function of the baryon number density $\rho_B/\rho_N^0$ at $T=20$ MeV. 
}
\label{fig:QH20}
\end{center}
\end{figure}
In the case of $T=20$ MeV, it is seen that the quark-hadron phase transition occurs at $\mu_B\approx1190$ MeV as may be seen from Fig.~\ref{fig:QH20} on the left.
Thus, the nuclear/hadron phase is realized in the region of $\mu_N<1190$ MeV and the quark phase in the region of $3\mu_q>1190$ MeV.
From the right side of Fig.~\ref{fig:QH20}, it is seen that the coexistence of nuclear and quark phases occurs from $\rho_N (=\rho_B) \sim 2.49\rho_N^0$ 
to $\rho_q \sim 9.99\rho_N^0$ ($\rho_B \sim 3.33\rho_N^0$) and a first-order chiral phase transition is realized there.
As compared with the case of $T=0$, it is seen that the crossing point (quark-hadron phase transition point) moves toward the lower left 
(small chemical potential and low pressure side) and the density region of the coexistence of nuclear and quark phases becomes smaller 
with increasing temperature from Fig.~\ref{fig:QH20}.

% Quark-Hadron(T=40)
\begin{figure}[t]
\begin{center}
   \begin{tabular}{ccc}
     \resizebox{70mm}{!}{\includegraphics{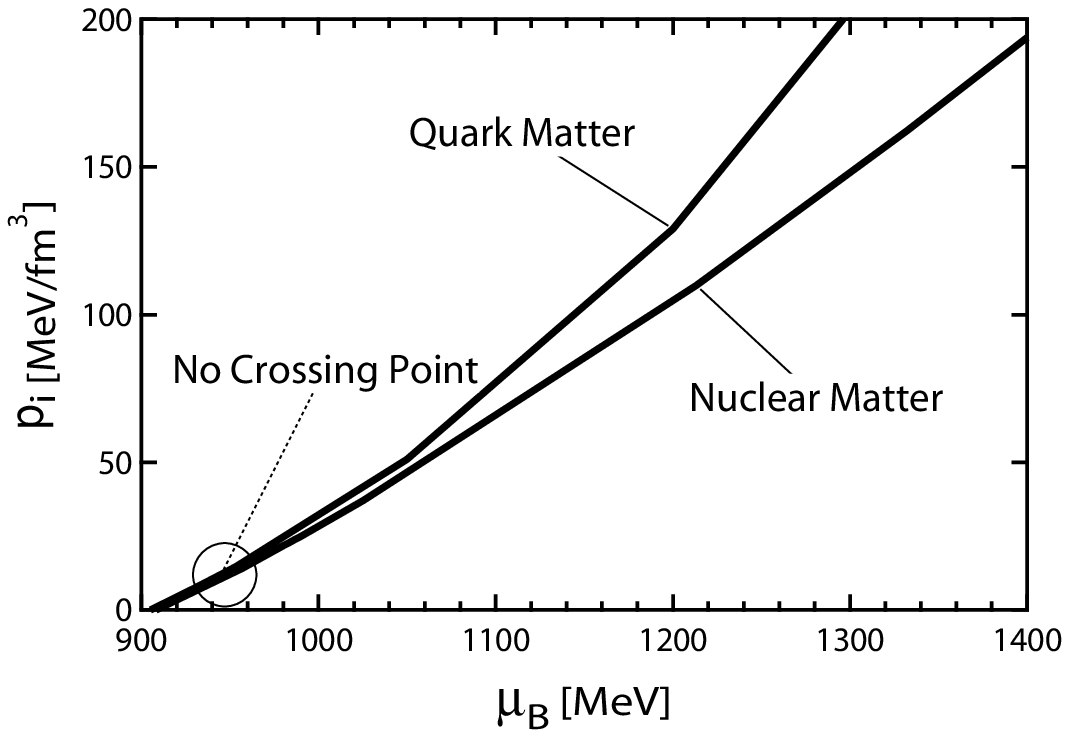}} &
     \resizebox{70mm}{!}{\includegraphics{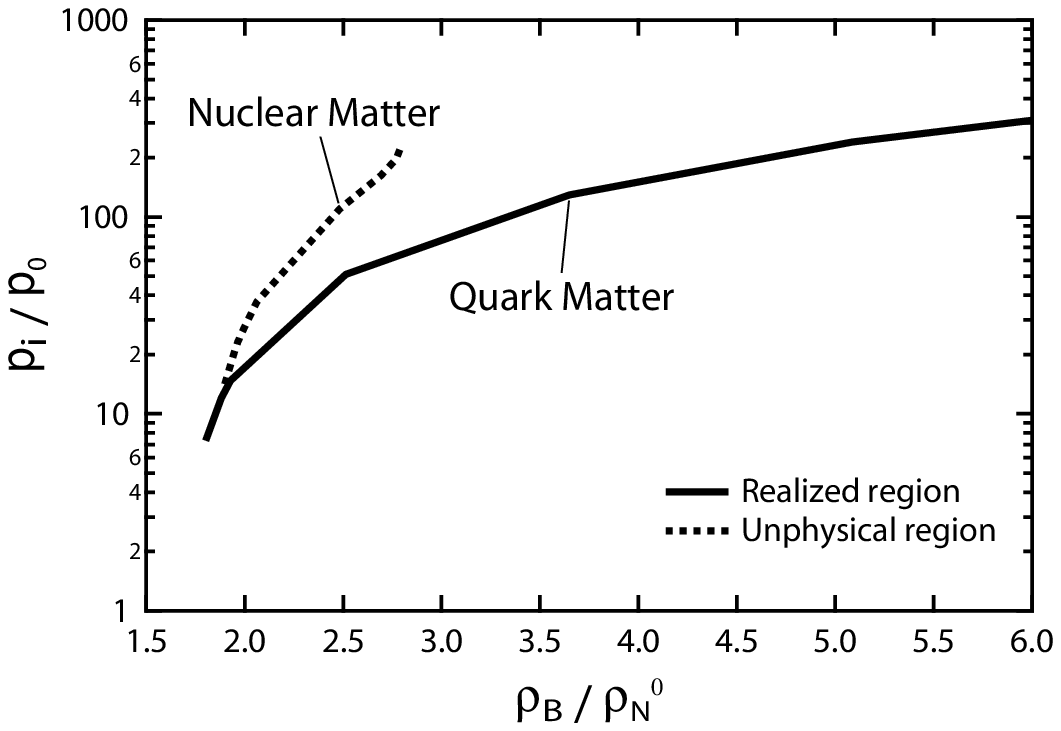}} 
   \end{tabular}
\caption{The first-order quark-hadron phase transition disappears at $T=40$ MeV in the extended NJL model with $G_{sv}^q\Lambda^8_q=-68.4$.
Left: The pressure $p_i$ as a function of the baryon chemical potential $\mu_B$ at $T=40$ MeV.
Right: The pressure presented on a logarithmic scale $p_i/p_0$ as a function of the baryon number density $\rho_B/\rho_N^0$ at $T=40$ MeV. 
}
\label{fig:QH40}
\end{center}
\end{figure}
In the process of increasing temperature, at a certain temperature, this crossing point vanishes.
This situation is depicted in Fig.~\ref{fig:QH40} with $T=40$ MeV.
In this figure, there is no crossing point which represents the transition point, so that this leads to the situation in which the first-order quark-hadron phase transition has already finished.
%Also, it is seen that the only quark phase is realized without the nuclear phase from Fig.~\ref{fig:QH40}.
%In this paper, however, we do not take account of this situation since this situation is beyond the applicability of our model.
%The phase transition is the second order as was mentioned in \S\S 3.1.

%%%%%%%%%%%%%%%%%%%%%%%%%%%%%
%%	  					%%
%%	      Phase diagram		%%
%%          					%%
%%%%%%%%%%%%%%%%%%%%%%%%%%%%%
\section{Phase diagram}\label{sec:4}
In this section, we present the phase diagram in the extended NJL model with the scalar-vector eight-point interaction at finite temperature and chemical potential.
Then, in the quark matter, we also investigate the effects of the scalar-vector coupling constant $G_{sv}^q$ 
which has influence on the chiral phase transition in the phase diagram.
Namely, we discuss how the chiral phase transition line is affected by varying the strength of the scalar-vector interaction.

%%  $G_{sv}^q=0$  %%
%Phase diagram with no scalar-vector interaction
\subsection{Phase diagram with $G_{sv}^q=0$}\label{subsec:4-1}
First, we present the phase diagram without scalar-vector interaction.
Thus, for the quark matter, we investigate the chiral and quark-hadron phase transition by using the original NJL model Lagrangian density without the scalar-vector interaction, that is, $G_{sv}^q=0$.

\begin{figure}[t]
\begin{center}
   \begin{tabular}{c}
     \resizebox{68mm}{!}{\includegraphics{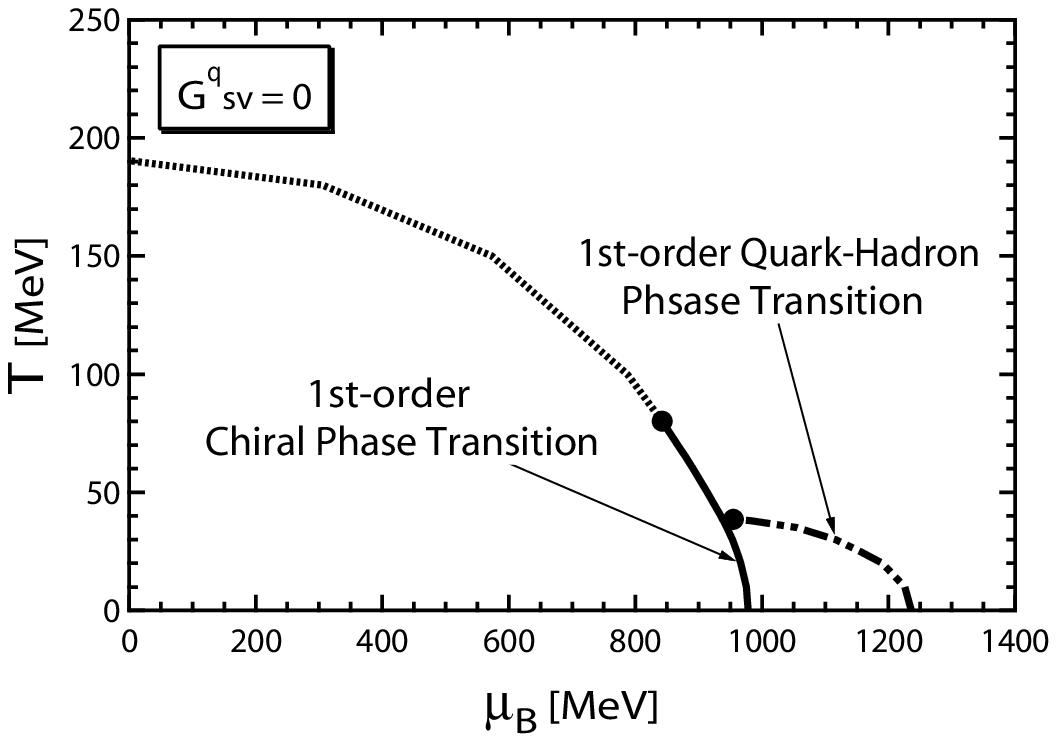}} \\
     \resizebox{68mm}{!}{\includegraphics{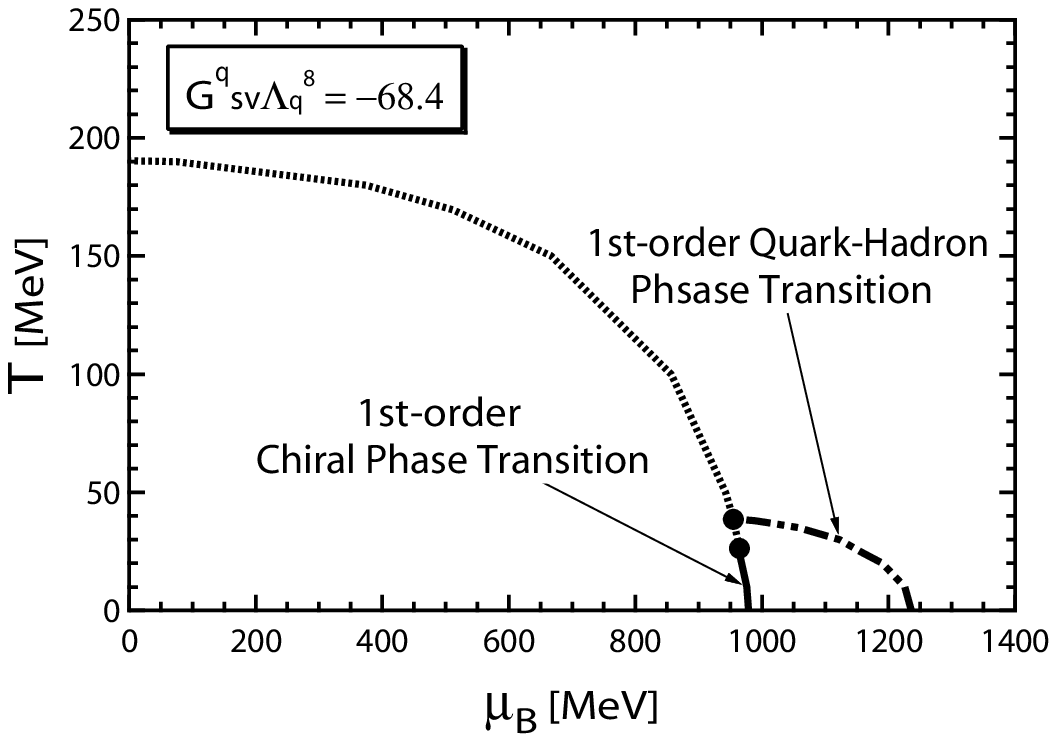}} \\
     \resizebox{68mm}{!}{\includegraphics{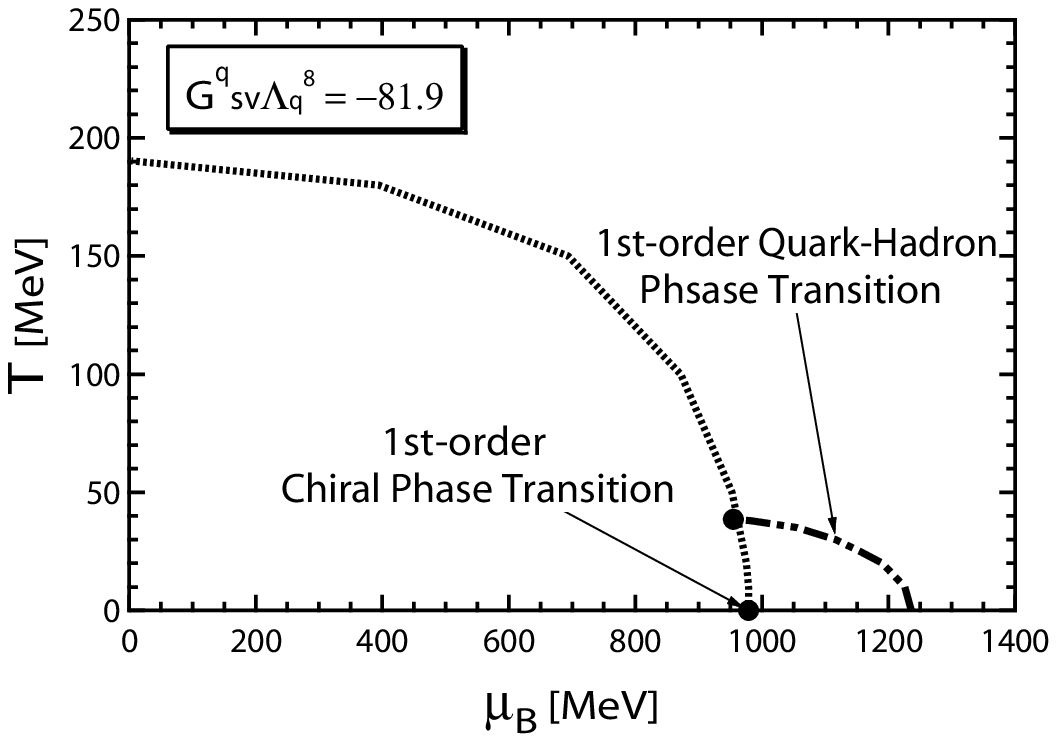}} 
   \end{tabular}
\caption{The phase diagrams in the extended NJL model with $G_{sv}^q$=0 (top), $G_{sv}^q\Lambda^8_q=-68.4$ (middle) and $G_{sv}^q\Lambda^8_q=-81.9$ (bottom) 
are depicted as a temperature-baryon chemical potential ($T$-$\mu_B$) plane.
The solid, dotted and dash-dotted curves indicate the 1st-order chiral phase transition, the 2nd-order 
chiral phase transition and the 1st-order quark-hadron phase transition, respectively.
The endpoint of the 1st-order chiral phase transition is at $(\mu_B, T)\simeq(842, 80)$ MeV (top), $(\mu_B, T)\simeq(964, 26)$ MeV (middle) 
and $(\mu_B, T)\simeq(979, 1)$ MeV (bottom).
For the 1st-order quark-hadron phase transition, the position of the endpoint is at $(\mu_B, T)\simeq(955, 39)$ MeV.}
\label{fig:pd}
\end{center}
\end{figure}

%fig.pd-0
Figure~\ref{fig:pd} on the top shows the phase diagram with $G_{sv}^q=0$ as a function of temperature and baryon chemical potential.
Here, the vertical and horizontal axes represent temperature $T$ and baryon chemical potential $\mu_B$ ($=\mu_N=3\mu_q$), respectively.
The solid curve represents the critical line of the first-order chiral phase transition and the dotted one represents that of the second-order chiral 
phase transition.
%The reason why we call the ``not first-order" chiral phase transition is that we can only be sure if the phase transition 
%is a first-order one as is mentioned before.
As may be seen from this figure, the critical line of the first-order chiral phase transition emerging from a point in the $T=0$ and $\mu_B\approx978$ MeV terminates at $(\mu_B, T)\simeq(842, 80)$ MeV.
Also, the critical temperature at vanishing chemical potential, $\mu_B=0$, is found to be about $190$ MeV.
The dash-dotted curve represents the first-order quark-hadron phase transition.
In this figure, it is seen that there is the endpoint of the first-order quark-hadron transition at $(\mu_B, T)\simeq(955, 39)$ MeV.

%
% phase diagram
%
%

%%  $G_{sv}^q\Lambda^8_q=-68.4$  %%
%Phase diagram with the scalar-vector interaction
\subsection{Phase diagram with $G^q_{sv}\Lambda^8_q=-68.4$}\label{subsec:4-2}
Next, we show the phase diagram with a scalar-vector interaction.
For the scalar-vector interaction strength, we consider the value $G_{sv}^q\Lambda^8_q=-68.4$ used in the previous section.

%fig.pd-68.4
Figure~\ref{fig:pd} in the middle shows the phase diagram in the $T$-$\mu_B$ plane with $G_{sv}^q \Lambda_q^8=-68.4$.
The critical line of the first-order chiral phase transition is depicted as a solid curve and that of the first-order quark-hadron phase transition as a dash-dotted curve.
The dotted curve represents the critical line of the second-order chiral phase transition.
This critical line of the first-order chiral phase transition emerges at $T=0$ and $\mu_B\approx 979$ MeV and terminates at $(\mu_B, T)\simeq(964, 26)$ MeV.
For the critical line of the first-order quark-hadron phase transition emerging from a point in the $T=0$ and $\mu_B\approx1236$ MeV, 
it terminates at $(\mu_B, T)\simeq(955, 39)$ MeV.
Then, the endpoint of the first-order quark-hadron phase transition is located on the critical line of the chiral phase transition as may be seen 
from Fig.~\ref{fig:pd} in the middle.

%chiral symmetric nuclear matter
Here, it should be noted that there is a region where the quark-hadron phase transition occurs after the chiral symmetry restoration in the nuclear phase side, in which the nucleon mass is zero.
% at the critical density of the first-order quark-hadron phase transition.
Namely, this suggests that a phase which is chiral symmetric but the elementary excitation from there is nucleonic (hadronic) could exist just before the phase transition from the nuclear phase to the quark one.
Recently, McLerran and Pisarski have proposed a new state of matter, the so-called 
quarkyonic matter\cite{MaC}, which is a phase characterized by chiral symmetry restoration and confinement based on large $N_c$ arguments.
Thus, this chiral symmetric nuclear phase predicted by our model may possibly correspond to the quarkyonic phase.

%%  $G_{sv}\Lambda^8_q=-81.9$  %%
\subsection{Phase diagram with $G^q_{sv}\Lambda^8_q=-81.9$}\label{subsec:4-3}
Finally, we show the phase diagram with the stronger scalar-vector interaction.
We set the value $G^q_{sv}\Lambda^8_q=-81.9$, in which $m_q(\rho_q/3=\rho^0_{N})=0.63m_q(\rho_q=0)$. 

%fig.pd-81.9
Figure~\ref{fig:pd} on the bottom shows the phase diagram in the $T$-$\mu_B$ plane with $G_{sv}^q\Lambda^8_q=-81.9$.
The critical line of the first-order chiral phase transition depicted as a solid curve emerges at $T=0$ and $\mu_B\approx 979$ MeV 
and terminates soon at $(\mu_B, T)\simeq(979, 1)$ MeV.
As for the critical line of the first-order quark-hadron phase transition depicted as a dash-dotted curve, the same diagram as previous two ones is obtained.

% Gsv-independence Q-H
The reason why the behavior of the first-order quark-hadron phase transition line seen in Fig.~\ref{fig:pd} does not change is 
that the critical line of the first-order quark-hadron phase transition exists in the chiral symmetric phase, 
in which $m_q=0$ and $\lal{\wb \psi}_q \psi_q\rar=0$, that is, the quark-hadron phase transition occurs after the chiral phase transition 
for the unstable quark phase with $G_{sv}^q=0$, $G_{sv}^q\Lambda_q^8=-68.4$ and $G_{sv}^q\Lambda_q^8=-81.9$.
Here, in the expression of the pressure $p_q$ in Eq.~(\ref{eq:2-16}), there is no $G_{sv}^q$-dependence since $\lal{\cal{H}}_{MF}^q\rar$ 
in Eq.~(\ref{eq:2-12}) and $\mu_q^r$ in Eq.~(\ref{eq:2-6}) do not depend on $G_{sv}^q$ due to $\lal{\wb \psi}_q \psi_q\rar=0$.

% Gsv-independence Chiral
From Fig.~\ref{fig:pd}, by varying the strength of $G_{sv}^q$, it is seen that the critical line of first-order chiral phase transition shrinks 
with increasing $G_{sv}^q$.
Namely, $G_{sv}^q$ acts to move the endpoint of the first-order chiral phase transition toward larger baryon chemical potentials and the lower temperatures.
In the case of $G_{sv}^q\Lambda^8_q=-68.4$, the endpoint of the first-order quark-hadron phase transition is located on the chiral phase transition line.
Also, the chiral symmetry restoration is shifted toward the larger baryon chemical potential.
In the case of the stronger $G_{sv}^q$, 
%that is, $G_{sv}^q\Lambda^8_q>-81.9$, this is beyond the applicability of the model for the chiral phase transition.
the line of the first-order chiral phase transition disappears.

%%%%%%%%%%%%%%%%%%%%%%%%%%%%%
%%	  					%%
%%	      	  Summary		%%
%%          					%%
%%%%%%%%%%%%%%%%%%%%%%%%%%%%%
\section{Summary and concluding remarks}
% objective+results
The quark-hadron phase transition at finite temperature and baryon chemical potential has been investigated following Ref~\citen{TPPY} in the extended NJL model with the scalar-vector eight-point interaction.
In this model, as a first attempt of investigation of the quark-hadron phase transition, the hadron side was regarded as a symmetric nuclear matter and the quark side as a free quark phase with no quark-pair correlation.
Here, the single nucleon in the nuclear matter and the single quark in the quark matter were treated as a fundamental fermions with $N_c^N=1$ and $N_c^q=3$, respectively.
Then, in this model, the nuclear saturation property has been well reproduced for the nuclear matter side.
On the other hand, for the quark matter side, there is one free parameter, $G_{sv}^q$.
This model parameter, in this paper, was not fixed since there is no criterion to determine the value of $G_{sv}^q$ by using physical quantities 
in this stage.
By introducing this parameter $G_{sv}^q$, the effective density-dependent coupling constant was obtained as $G_s^q(\rho_q)=G_s^q (1-G_{sv}^q/G_s^q\cdot\rho_q^2)$.
Hence, the $G_{sv}^q$ term plays an important role in pushing the chiral symmetry restoration point to higher density side for the quark matter.
Thus, the parameter $G_{sv}^q$ controls the chiral symmetry restoration point and/or the strength of the partial chiral symmetry restoration in the nuclear medium. 
%Incidentally, for nuclear matter, the scalar-vector attractive interaction $G_{sv}^N$ plays the role of lowering the incompressibility. 
As for the description of the quark-hadron phase transition, we have calculated the pressure of the nuclear matter and 
the quark matter, and determined the realized phase by comparing their pressures.
As a result, a first-order quark-hadron phase transition is obtained at finite temperature and baryon chemical potential.
Here, the end point of the first-order quark-hadron phase transition is at $(\mu_B, T)\simeq(955, 39)$ MeV with $G_{sv}^q\Lambda_q^8=-68.4$.
This phase boundary is not changed even by varying the strength of the scalar-vector interaction because of $G_{sv}^q$-independence.
As for the effects of the scalar-vector coupling constant $G_{sv}^q$ on the chiral phase transition, the critical line 
of the first-order chiral phase transition shrinks with increasing $G_{sv}^q$.
Namely, $G_{sv}^q$ acts to move the endpoint of the first-order chiral phase transition toward a larger $\mu_B$ and a lower $T$.
%Here, in this paper, 
%we concentrate on the first-order phase transition. 
%To analyze order of other phase transition is a future problem.
%we only consider the first-order phase transition since there is no method for determining other order of phase transition.
%This is a future problem.

%quarkyonic phase
From the phase diagram in Fig.~\ref{fig:pd}, it should be noted that there is an interesting phase where the quark-hadron phase transition occurs after the chiral symmetry restoration in the nuclear matter.
%This might appear as an exotic nuclear phase in which chiral symmetry is restored but the elementary excitation modes of matter are nucleonic.
This might appear as an exotic phase which is the nuclear phase, not quark phase, while the chiral symmetry is restored in terms of the quark matter.
This phase may possibly correspond to the quarkyonic phase\cite{MaC} which is introduced as a chiral symmetric confined matter.

%CSC+neutron star
In this paper, we have ignored the color superconducting phase\cite{CSC}.
However, this phase may exist at finite density system.
Thus, a possible next challenging task is to investigate the phases of nuclear matter including the nuclear superfluidity 
and quark matter including the color superconducting state, that is,
the nucleon pairing on the side of nuclear phase and the quark pairing on the side of quark phase. 
% should be taken into account at finite temperature and density.
Further, it is widely believed that neutron star matter undergoes a phase transition to quark matter at high temperature and/or density.
Thus, it is also interesting to investigate the phase transition between neutron star matter and quark matter.
This leads to the understanding and development of the physics of neutron star.

\section*{Acknowledgement} 
One of the authors (T.-G. L.) would like to express his sincere thanks to Professor K. Iida, Dr. E. Nakano, Dr. T. Saito, Dr. K. Ishiguro 
and the members of Many-Body Theory Group of Kochi University for valuable comments and fruitful discussions. 
One of the authors
(J. P.) acknowledges valuable discussions with Steven Moszkowski.
One of the authors (Y. T.) is partially supported by the Grants-in-Aid of the Scientific Research 
(No.23540311) from the Ministry of Education, Culture, Sports, Science and 
Technology in Japan. 

%\appendix
%\section{}

%\appendix
%\section{}

\end{document}